# The Impact of Knowledge of the Issue of Identification and Authentication on the Information Security of Adolescents in the Virtual Space

LJERKA LUIĆ, DRAŽENKA ŠVELEC-JURIČIĆ, PETAR MIŠEVIĆ
Media and Communication
University North
Trg dr. Žarka Dolinara 1, Koprivnica
CROATIA

*Abstract:* - Information security in the context of digital literacy is a digital skill that enables safe and purposeful movement through virtual space. Due to rapid and unstoppable technological progress, multiplying opportunities and pushing the boundaries of digital technology and the Internet, the interest of the state and institutions within the state is to raise digital competencies of citizens, with special emphasis on children and youth as the most vulnerable groups of Internet users. The age limit and frequency of use of the Internet by young generations has been moved back a year due to the COVID-19 pandemic, and the concern for information security of young people is increasingly emphasized. If, and to what extent, knowledge of the issue of identification and authentication affects the information security of high school students aged 16 to 19 in the virtual space, the research question addressed by the authors of this paper was to determine which student behaviors pose a potential danger compromising their information security by establishing a correlation between the variables that determine student behavior and the variables used to examine their level of security in a virtual environment. The research was conducted using a questionnaire on a sample of high school students in the Republic of Croatia, the results of which showed that some students practice behaviors that are potentially dangerous, make them vulnerable and easy targets of cyber predators and attackers, which is why there is cause for concern and a need for a additional education of children of primary and secondary school age in the field of information security in the form of the introduction of the subject Digital Literacy. Based on the results, a model for assessing the level of digital literacy of adolescents that affect information literacy can be designed, but also further related research in the field of information literacy of children and youth can be conducted.

*Key-Words:* - Identification, Authentication, Digital Literacy, Information Security, Knowledge, Adolescents

Received: July 9, 2021.  Revised: September 24, 2021.  Accepted: September 27, 2021.  Published: October 1, 2021.

## 1 Introduction

In today's fast-growing and rapidly evolving digital environment where reality is increasingly being given the prefix virtual, the aspect of information security is becoming imperative in all areas of human action affected by digital technology. In recent years we have witnessed major cyber-attacks all over the world, bringing information security to importance in the public space while at the same time raising awareness of the strong dependence on digital technology. [1] The ubiquitous information and communication environment and awareness of it at the global level generates the necessity of preparedness for digital threats and becomes extremely important for all organizational systems. Although it is not possible to completely eliminate all threats in the digital space, by raising digital education, and thus risk identification and management, appropriate techniques and processes, it is possible to raise the competence of systems to become resistant to digital threats. [2]. With the appearance of smartphones, social networks and mobile applications, the flow of information has become incomparably faster and easier, bringing with it the benefits and numerous opportunities for improvement in the field of business and private life. At the same time, new digital technology, due to its limitlessness, raises issues related to security, and privacy, data protection and data abuse are becoming unavoidable concepts in all spheres of social life. [3]

"National Information Security Program in the Republic of Croatia" [4] is linked to the guidelines of the European Union's e-Europe 2005 Program, the e-South East Europe Initiative (eSEE Initiative) under the auspices of the Stability Pact and the NATO Membership Action Plan, and defines the goals of information security, competencies and activities of institutions operating in in the field of information security and the method of coordination of information security factors in the Republic of Croatia. The program envisages the introduction of information security measures in all segments of society over a period of four years.

The program starts from the definition of information as data with a certain meaning, ie knowledge that can be transmitted in any form (written, audio, visual, electronic or otherwise), while information security is determined by maintaining confidentiality, integrity and availability. The interpretation of these three aspects depends on the context in which they arise, but also the needs of members of a particular group and the rules that determine behaviors in the digital environment. [5] Confidentiality refers to the security of the system that information will be available only to those who have authorized access. Security in this aspect is achieved by user identification and authentication thus the system restricts access to unauthorized users. Integrity is the area of security by which the system defends itself and resists possible attacks in such a way that users are prevented from unauthorized modification, deletion or destruction of content. The third aspect is availability, which implies the availability of the service to authorized users and its role is to ensure the availability of information and content to users who have authorized access. [6]

In accordance with the change of the way and channels of communication, information is becoming the most valuable and most important currency, and therefore the issue of information security in all countries is perceived at the level of the state and the law. As part of the development of digital competencies provided for the Council of Europe by the document "The Digital Competence Framework for Citizens DigComp 2.1." one of the 5 dimensions of digital competencies is security. This area includes the protection of devices, the protection of personal data and privacy, the protection of health and well-being, and the protection of the environment. [7] Based on the document of the Council of Europe, national strategies and policies of EU member states on raising the level of digital competencies of citizens are created, thus Austrian Framework "Digital Competence Framework for Austria DigComp 2.2." points out that nothing proposed by the document can happen by accident, but must be introduced systematically, focused on objectives and sustainably, including a range of measures such as the establishment of appropriate institutions to implement activities to improve digital competencies that can meet challenges. [8]

In addition to European documents, for free use there are tools for self-assessment of the level of digital competencies which examine the level of 21 digital competences provided for in the Council of Europe document, and there are also extended versions (Austrian version of the self-assessment questionnaire) where the level of digital competences is tested in 6 areas through 3 levels. Elements of available questionnaires that were partly used in creating the questionnaire for the purposes of this research.

Given that digital competencies, digital skills, Internet security and information security are inseparable and highly contextual concepts, it is necessary to develop them simultaneously and with the same dynamics constantly complementing and following the development of technology and the scope of digital space.

Information security must be built into systems and processes by understanding and perceiving possible threats and risks, and it cannot be thought of only as a system of protection, setting barriers or setting a secure password. In any system or process, the weak points of information security do not arise from technological imperfections, but it is a person who, despite his knowledge and competencies, endangers his own information security with his actions. [1]

Information security is user-oriented, and the basic future module of increasing the security framework should be based on the identification and authentication of users of certain applications, assuming that the system recognizes the malicious person trying to log in. The identification and authentication process are an integral part of the interface of almost every application used, with some data to be accessed requiring a strong password composed of uppercase and lowercase letters, numbers and special characters. [11]

Of great importance is the speed and efficiency of resolving incidents related to information security of users. Cyber-attacks often cause personal and business data to be compromised, and are caused by viruses, worms, Trojan horses, spyware, or similar programs that can cause harm to Internet users. [12]

## 2 Related Work

Based on Council of Europe documents, in order to improve the digital competences of civil citizens and educational curricula, European countries define national policies for the development of digital competences in which activities to improve digital competences for children and young people are recognized as the most vulnerable group.

A study conducted in 2011 involving 25,242 children aged 9 to 16 from 25 European countries showed that the average age of first internet access was 9 years at the level of all European countries covered by the study, in Denmark and Sweden 7 years, and in other northern countries (Norway, Finland, the Netherlands and the United Kingdom) 8 years.

In terms of Internet access frequency, 66% of children are online every day and 33% once or more a week, while on average they spend 88 minutes a day online. [9] Such data are certainly worrying because the trend of using computers and mobile devices and accessing the Internet is becoming more frequent and intense, especially in the last year due to the COVID-19 pandemic.

Considering that school education curricula are not changing at the same rate as technology advances and the real need for children to access the Internet, the question arises as to whether students are sufficiently educated in the field of information security in the virtual space.

Adolescence is the most sensitive period of growing up of a human being when it is important for each of them to be accepted in the real world, and especially in a virtual environment whose premise is the presence on the Internet and developed digital communication. [10]

## 3 Methodology

This paper starts from the assumption that high school students aged 16 to 19 endanger their information security by their behaviors in the online environment.

The aim of the research was to determine which student behaviors pose a potential threat to information security, focusing on the area of personal data protection through identification and authentication as the first line of defense in raising the level of information space security.

Starting from the problem and the aim of the research, the research question was asked whether the knowledge of the issue of identification and authentication affects the information security of high school students aged 16 to 19 in the virtual space.

The research was conducted using a survey questionnaire on a sample of high school students in the Republic of Croatia, aged 16 to 19 years. The questionnaire explored the habits of high school students regarding information security and data protection in a virtual environment, observing their activities regardless of the context of school dynamics. In the introductory part, along with the identification of the researcher, the purpose and aim of the research are presented.

The survey contains 22 questions, of which three are questions of socio-demographic structure (gender, grade, type of high school). Closed-ended questions are a combination of multiple choice, dichotomous, and Likert-based 5-point scale, with a value of 1 (one) being the lowest value and 5 (five) being the highest.

The survey was created in the form of a Google form. The research was conducted in the period from 7 to 12 June 2021. The link to the survey was sent to high school students of the Republic of Croatia in groups on social networks. Group members sent links to other potential respondents using digital communication channels.

The data obtained from the research are shown in the tables in frequencies (f) and percentages (%). A descriptive-statistical method and graphical presentation of data were used to present the processed research results. Methods of induction, deduction, analysis and synthesis were used in drawing conclusions.

## 4 Results

The survey was completed by 340 high school students in the Republic of Croatia, of which 75% were women and 25% were men. Of the total number of respondents, most (49%) attend the 4th grade of high school (18 and 19 years of age), most high school students who completed the survey attend the gymnasium program (65%).

### 4.1 Password Strength for Accessing Services and Applications used by Students

When asked what password they use to access services and applications, more than half of the respondents (56%) answered that they use a 10-character password that is a combination of letters, numbers and special characters, while 12% use a password that consists of more than 10 characters a combination of letters, numbers, and special characters. More than a third of students (32%) use simple passwords that provide weak protection such as a personal name from the family environment, consecutive single-digit numbers or a word.

Table 1. Password Strength for Accessing Services and Applications used by Students

|  | f | % |
|---|---|---|
| a 10-character password with a combination of letters, numbers, and special characters | 191 | 56,2 |
| a password of more than 10 characters with a combination of letters, numbers, and special characters | 41 | 12,1 |
| my name and/or date of birth or members of my family | 51 | 15,0 |
| word in Croatian or English | 44 | 12,9 |
| consecutive one-digit numbers | 13 | 3,8 |
| Total | 340 | 100,0 |

### 4.2 Password Storage Security Levels

When it comes to dealing with forgetting the password, 54% of students answer that in that case they will click on "Forgot password" to get a new password in the mail, which implies that the passwords are not stored anywhere, but they remember them. 21% of students have all their passwords written down on paper, while 14% of them say that they cannot forget their password because it is related to their personal data. The most insecure way to store a password by saving it to a computer folder is used by 9% of students.

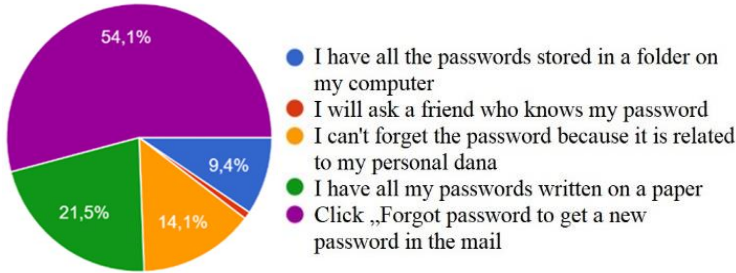

Fig. 1: Password Storage Security Levels

### 4.3 Security Levels of using Passwords to Access Internet Services

When asked if they use different passwords to access different services and applications, most respondents (52%) answered that they have several passwords that they use to access all services and applications. Almost a third of students (28%) use one password to access all services and applications, while the most secure way of behaving, accessing each service or application with a different password, is used by only 20% of respondents.

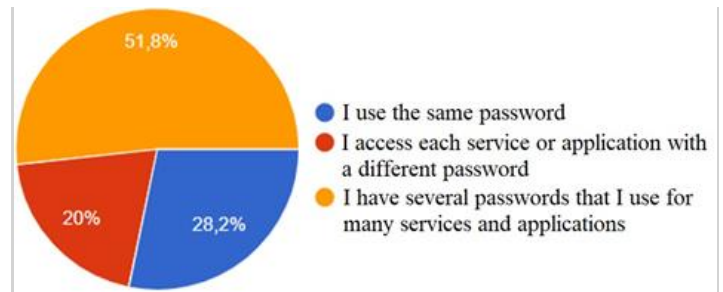

Fig. 2: Security Levels of using Passwords to access internet services

### 4.4 Password Change Frequency

According to the statements of the respondents about the frequency of changing the password, 56% of them say that they change the password when they suspect the possibility of abuse. Almost a third of respondents (29%) do not change their password once while using a particular application or service. A small proportion of respondents (15%) change their password at intervals of every few months or once a year.

Table 2. Password Change Frequency

|  | f | % |
|---|---|---|
| I change my password when I suspect a possible misuse | 192 | 28,8 |
| I use the password I create as long as I use that application | 98 | 12,1 |
| about once a year | 32 | 9,4 |
| every few months | 18 | 5,3 |
| Total | 340 | 100,0 |

### 4.5 Level of Protection of the Device From Unwanted Access

The results presented in the following table provide data on the respondents' responses to take protective actions in order to prevent unwanted access to digital devices. When asked what they would do if they leave their device in a room where they are familiar people, most of them answer (74%) that all their devices are password protected so that no one can access them. A smaller percentage (14%), but not negligible, states that they leave the device turned on because they have confidence in their friends and acquaintances. Only 8% of respondents state that, for security reasons, they will turn off the device so that no one can access their device and data.

Table 3. Level of Protection of the Device from Unwanted Access

|  | f | % |
|---|---|---|
| all my devices are password protected | 253 | 74,4 |
| I leave the device on because I trust my friends/acquaintances | 47 | 13,8 |
| I turn off the device so no one can access it | 22 | 8,5 |
| I put the device in sleep mode | 11 | 3,2 |
| Total | 340 | 100,0 |

### 4.6 Checking Email Compromise

When asked if they check the compromise of their email address on the "Have I been pwned?" service almost half of the respondents (45%) answer that they do not know about the existence of such a way of checking the security of the email address, while 23% of the respondents answer that they do not check for possible compromise of the email address. A small proportion of respondents (14%) state that they check only when they suspect a possible compromise of their email address, and 11% check when they receive an email address that seems dangerous or harmful to them. Of the total number of high school students who participated in the survey, only 6% regularly check for possible compromise of the email address.

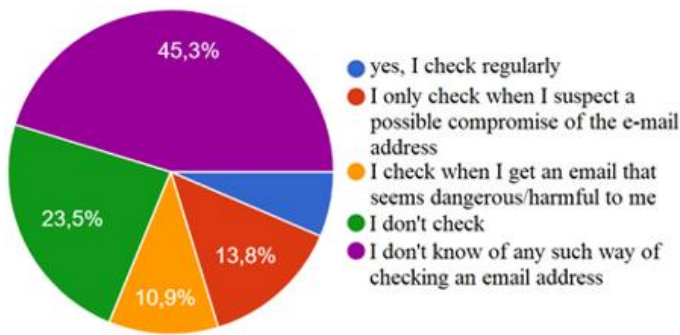

Fig. 3: Checking Email Compromise

Analyzing the results shown in Fig. 3, it can be concluded that most of the high school students who participated in the research were not victims of identity theft or any form of data misuse. A small proportion of respondents (22%) state that they are not sure whether they have been victims of personal data theft on the Internet, while 13% of students state that they have been victims of personal data or identity theft or some other form of data misuse.

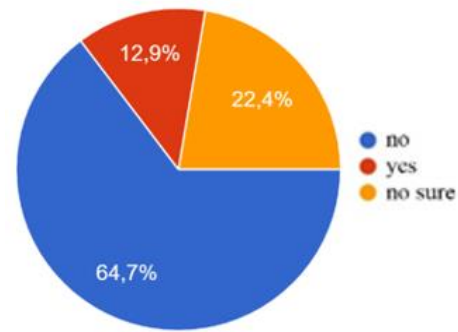

Fig. 4: Proportion of victims of personal data theft.

## 5 Discussion

The starting point of the research was the research question whether the knowledge of identification and authentication affects the information security of adolescents in the virtual space, which was obtained by analyzing the results of the research on a sample of 340 respondents. The analysis of the research results links the knowledge and application of the principles of identification and authentication in order to increase security in the virtual space with the information security of high school students aged 16 to 19 in a way that their online behavior determines their level of information security.

More than a third of respondents use simple passwords that provide poor protection and are easy to detect because they are related to students' personal data (usually personal names) or consist of consecutive numbers or some frequent words, making them vulnerable in the virtual space and exposed to potential threats and cyber-attacks. As for the level of password storage security, almost a third of students experience potentially dangerous behaviors through storing a password by saving it in a computer folder or creating passwords that are associated with their personal names and therefore do not write them down.

The level of user security when accessing Internet services is also determined by creating a new password for each new Internet service. Using the same password to access all existing accounts and services compromises the security of users on the Internet. Using the same password to access, for example, an email address and a bank account makes us vulnerable and potential victims of personal data theft and misuse. Almost a third of students (28%) use a single password to access all services and applications, which endangers the security of their personal data in the virtual space. Vulnerability of one third of respondents is visible through long-term use of the password without a tendency for the password to be changed at certain set time intervals or when there is a suspicion of data misuse.

One of the simplest and most reliable ways to prevent potential vulnerability in the virtual space is to regularly check the compromise of the email address on the "Have I been pwned?" whose existence is not known to as many as 45% of respondents. A quarter of respondents (25%) check the compromised e-mail address only when there is a suspicion of possible vulnerability, almost as many (23%) do not check at all, while a worryingly small percentage of students (6%) regularly check the possible compromise of their e-mail address.

Ultimately, the results show that 13% of the high school students who participated in the survey were either victims of identity theft or some other form of data misuse. Given the results that indicate that at least one third of students in virtual space practice behaviors that are potentially dangerous, make them vulnerable and easily accessible to victims of some form of cyberbullying, it can be concluded that the percentage of students who have experienced some form of violence in cyberspace is higher, as some of them state that they are not sure whether they have experienced some form of data misuse.

# 6 Conclusion

All elements of the aspect of confidentiality in the context of information security examined by this research, such as knowledge of the basic principles of implementation of identification and authentication processes in everyday use of digital devices in virtual space, are an integral part of school curricula in primary and secondary school in Croatia. Therefore, to a lesser extent we can talk about ignorance of how to protect against unwanted actions aimed at harming users and given the fact that the survey was completed by 65% of students attending high school and 49% of them aged 18 and 19 years.

The results of this research can be observed through the prism of non-application of what has been learned, thus endangering adolescents' own safe and efficient movement through digital space. Given that security skills in the use of virtual space and digital literacy are highly contextual concepts, such results suggest the need for additional education of primary and secondary school children in terms of strengthening digital competencies, including the domain of security through the introduction of Digital Literacy in primary education in correlation with the already existing subject Informatics.

Digital devices are an integral part of growing up and the lives of younger generations and they are the cause of a different, faster, multitasking way of thinking and functioning of generations of digital natives.

Adolescents belong to the generation of digital natives who were born and raised under the ubiquitous influence of the Internet and modern technologies, to whom multitasking is a natural behavior and who simultaneously use multiple sources of information and simultaneously receive and process information presented by text, image, sound and movement. In order to survive in such a technologically advanced world, digital natives live in both worlds (virtual and real) with the same intensity, integrating the experiences they gain with online and offline stimuli. The reason why young people do not use their knowledge about the dangers of Internet world can be found in their intention to solve the situation faster and easier, or in the insufficient awareness of the consequences of possible cyber-attacks.

The results of this research could be of great importance for teachers, principals and authors of educational curricula who will make their decisions based on awareness of the dangers to which young people are exposed in the online world.

The paper represents a significant contribution to the field of information security by identifying harmful and threatening actions and behaviors of adolescents during the process of identification and authentication based on which it is possible to design a model for assessing the level of digital literacy affecting information security.

Based on the results of this research, it is possible to conduct related research in the field of information security of primary school children or research on information security of adolescents in other European countries in order to obtain a broader and comparative picture.

**Contribution of Individual Authors to the Creation of a Scientific Article (Ghostwriting Policy)**
Ljerka Luić has managed and coordinated research activity planning and execution.
Draženka Švelec-Juričić has created questionnaire, developed methodology and conducted research process.
Petar Mišević was responsible for the idea and research goals.

**Sources of Funding for Research Presented in a Scientific Article or Scientific Article Itself**
The publication of this paper was possible by the grant from the University North.